# Electro-Optic Effects in Colloidal Dispersion of Metal Nano-Rods in Dielectric Fluid

**Andrii B. Golovin [1], Jie Xiang [1,2], Heung-Shik Park [1,2], Luana Tortora[1], Yuriy A. Nastishin [1,3], Sergij V. Shiyanovskii [1], and Oleg D. Lavrentovich [1,2],***

1 Liquid Crystal Institute, Kent State University, Kent, Ohio, 44242
2 Chemical Physics Interdisciplinary Program, Kent State University, Kent, Ohio, 44242
3 Institute of Physical Optics, 23 Dragomanov Str. Lviv, 79005, Ukraine

E-Mails: agolovin@kent.edu (A. B. G.); jxiang@kent.edu (J. X.); hpark3@kent.edu (H.-S. P.); ltortor1@kent.edu (L. T.); ynastish@kent.edu (Y. A. N.), sshiyano@kent.edu (S. V. S.), olavrent@kent.edu (O. D. L.)

* Author to whom correspondence should be addressed; E-Mail: olavrent@kent.edu;
  Tel.: +1-330-672-4844; Fax: +1-330-672-2796.

**Abstract:** In modern transformation optics, one explores metamaterials with properties that vary from point to point in space and time, suitable for applications in devices such as an "optical invisibility cloak" and an "optical black hole". We propose an approach to construct spatially varying and switchable metamaterials that are based on colloidal dispersions of metal nano-rods (NRs) in dielectric fluids, in which dielectrophoretic forces, originating in the electric field gradients, create spatially varying configurations of aligned NRs. The electric field controls orientation and concentration of NRs and thus modulates the optical properties of the medium. Using gold (Au) NRs dispersed in toluene, we demonstrate electrically induced change in refractive index on the order of 0.1.

## 1. Introduction

Optical metamaterials represent artificial composites with building blocks of metal and dielectric nature intertwined at a sub-wavelength scale. When properly arranged, these building units lead to fascinating optical effects, such as negative refraction and sub-wavelength imaging. Optical metamaterials in which the electric permittivity and magnetic permeability vary in space and can be switchable, are of special interest. The reason is simple: by controlling the spatial variation of permittivity and permeability, one controls the local refractive index and thus the pathway of light in the medium. According to the Ferma's principle of least time, a light ray propagating from a point A to a point B follows a path that minimizes the travel time. For a small path element, the quantity to minimize is simply a product of the geometrical path length and the refractive index. Thus the spatially varying refractive index can make the light rays to follow curved trajectories. If these trajectories are designed to avoid a certain region of the medium, one obtains an invisibility cloak, as any object placed within this region would not interact with light [1, 2]. Potential applications of metamaterials with spatially varying properties are much wider than cloaking and extend from perfect magnifying lenses with sub-wavelength resolution [3] to optical "black hole" collectors [4, 5], as reviewed recently by Wegener and Linden [6]. To find the pathway of light theoretically, one uses the equivalence of coordinate transformations and renormalization of permittivity and permeability; this is why the field of study is called the "transformation optics" [1, 2, 7-9].

The fact that light rays follow curved trajectories in a medium with a varying refractive index has been known for a very long time in the physics of liquid crystals. In the simplest liquid crystal, the so-called uniaxial nematic, rod-like molecules align parallel to each other, along the common "director" $\hat{\boldsymbol{n}}$. The director is a unit vector with a property $\hat{\boldsymbol{n}} = -\hat{\boldsymbol{n}}$ (the medium is non-polar); it is also a local optic axis. The associated birefringence $\Delta n = n_e - n_o$ of a typical nematic formed by low-molecular-weight organic molecules is significant: the ordinary refractive index $n_o$ is often about 1.5, while the extraordinary index $n_e$ is about 1.7. In liquid crystals, the local orientation of molecules and thus the local optic axis can be made varying in space and time, for example, by setting proper surface alignment at the boundaries and applying an electric field to realign $\hat{\boldsymbol{n}}$ (a phenomenon at the heart of modern liquid crystal displays). The early liquid crystalline example for transformation optics has been presented by Grandjean in 1919 [10] (for a textbook reproduction, see Ref. [11]). Grandjean considered a cylindrical nematic sample in which the director was arranged radially. When such a structure is illuminated with light polarized normally to the axis of cylinder, the rays are bent away from the central axis and leave a segment of an opening angle $2\pi(1 - n_o/n_e)$ un-illuminated [10, 11]. This particular example represents, loosely speaking, half a cloak, as the trajectories are diverging. The limitation of a regular liquid crystal is that although the ellipsoid of refractive indices is changing its orientation in space, it cannot be shrunk or expanded at will.

In a metamaterial, the refractive index (or indices) can be made changing from point to point. An excellent example is the optical cloak proposed by the Shalaev's group [12]. A cylindrical shell of a (rigid) dielectric is penetrated with radial metal nano-wires. The metal filling factor increases as one moves from the outer to the inner surface of shell. The optic axis configuration is identical to the Grandjean's model, but in the metamaterial, $n_e$ changes with the radial coordinate, down to zero at the inner surface, while in the Grandjean's liquid crystal, $n_e = const$. The light trajectories in the cloaking shell first diverge and then converge, to restore a flat front as they pass around the shell [12]. By properly adjusting the radial variation of the refractive index, one can greatly reduce the visibility of an object enclosed by the shell [12]. Such a proper adjustment requires one to distribute small (sub-wavelength) elements in an essentially gradient manner, which represents a major technological difficulty [13-21]. Nowadays, metamaterials are fabricated by electron beam lithography, focused ion-beam milling [13], deposition of alternating metal and dielectric layers [16], or by growing metallic wires from within a dielectric medium [17]. These metamaterial structures should be more properly called "metasurfaces" or "metafilms" as their functionality is determined by only one or a few layers normal to the direction of propagating light [6]. For complex architectures, involving property variations along the three spatial dimensions and switching, new approaches are needed. Among the potential candidates are bottom-up self-assembly [18], alignment of NRs by a uniform electric field [19] or assembly through a non-uniform electric field [20].

Recently, we proposed that the next wave of metamaterials with spatially varying and even switchable optical properties can be based on dispersions of small (sub-wavelength) metal nanorods (NRs) in a dielectric fluid, controlled by a nonuniform electric field [22]. The gradients of the electric field pull the highly polarizable NRs towards the strongest field and also align them along the field lines. The reason is that the field-induced dipole polarization experiences different pulling force at the two ends of the NRs when the field is non-uniform. The effect is known as dielecrophoresis [20]. If the electric field is radial, for example, created by two concentric cylindrical electrodes, then the NRs align radially and condense near the inner electrode [22]. The structure is similar to the cylindrical cloak proposed in [12], with that difference that the location and orientation of NRs is determined by the dielectrophoretic forces and interactions between the NRs rather than by mechanical means. We used NRs that are much smaller than the wavelength of light, of length about (40-70) nm, to reduce light scattering. Previously, dielectrophoretic manipulation has been demonstrated for much larger supra-micron metal wires [23-28], but the viability of downscaling is not obvious, as the dielectrophoretic force acting on the particle is proportional to its volume [20] and might be too small at nanoscales. Similarly small NRs were previously studied under the action of a uniform electric field that can impose an orienting torque on the NRs [19, 29, 30]. In this work, we expand the scope of the original experiments [22], present new data for different dispersions of NRs and analyze the field-induced pattern analytically and numerically, in order to obtain information about the dielectrophoretic forces acting on NRs, field-induced spatial distribution of NRs, field-modified refractive indices and coefficients of absorption.

## 2. Experimental Materials and Methods

### 2.1. Dispersions of NRs in Toluene

We used dispersions of gold (Au) NRs in dielectric fluids, such as toluene and water. Au NRs can be grown by the so-called seed mediated process in water solutions of a cationic surfactant cetyltrimethylammonium bromide $(C_{16}H_{33})N(CH_3)_3Br$, abbreviated as CTAB. CTAB forms a charged bilayer around the NRs, preventing them from aggregation. When Au NRs reach the desirable length $l_{NR}$ and diameter $d_{NR}$, the NR dispersion is centrifuged and redispersed in deionized water. We also used water dispersions of Au NRs commercially available from Nanopartz, Inc.

For optical experiments with NR dispersions in glass containers, it is convenient to match the refractive index of the dispersive medium with the refractive index of glass. We use toluene with $n_t = 1.497$ measured at $\lambda = 589.3$ nm and 293 K; $n_t$ is close to the refractive index of borosilicate glass. To transfer Au NRs from water into toluene, we followed the approach developed by N. Kotov and P. Palffy-Muhoray groups, in which the Au NRs are functionalized with thiol terminated polystyrene [29, 30]. A 2 wt% solution of thiol terminated polystyrene (molecular weight 53,000, purchased from Polymer Source, Inc) in tetrahydrofuran is added by rapid stirring to the water dispersion of CTAB-stabilized Au NRs. The mixture is incubated overnight for hydrophobization-induced precipitation of NRs. The rods are collected after supernatant removal, by re-dissolution in toluene. The volume fraction of Au NRs in toluene was increased by centrifuging to $\eta_o \approx (4-8) \cdot 10^{-4}$. The typical volume fraction of Au NRs in water dispersions produced by Nanopartz, Inc. was $\eta_{water} = 7.4 \cdot 10^{-6}$.

To facilitate the study of spatial structure and optical properties, we use three types of dispersions: (1) "long/thin" NRs in toluene, with an average length $l_{NR}$=70 nm and diameter $d_{NR}$=12 nm, showing a longitudinal plasmonic absorption peak at $\lambda \approx 1040$ nm; (2) "short/thick" NRs in toluene, with $l_{NR}$=50 nm, $d_{NR}$=20 nm, and the absorption maximum at 725 nm; (3) "short/thin" NRs in water and toluene with $l_{NR}$=45 nm, $d_{NR}$=10 nm. The spectral properties of dispersions strongly depend on the dispersive medium and NRs geometry, in particular, on the aspect ratio $l_{NR}/d_{NR}$, Fig.1(b,c).

**Figure 1.** Transmission electron microscope image of "long/thin" Au NRs **(a)**, absorption spectra of toluene dispersions of "long/thin" and "short/thick" Au NRs **(b)**, and water dispersion of "short/thin" Au NRs at volume fraction $\eta_{water}$ (blue line), as well as toluene dispersions with volume fractions $4\eta_{water}$ (red line) and $50\eta_{water}$ (black line) **(c)**.

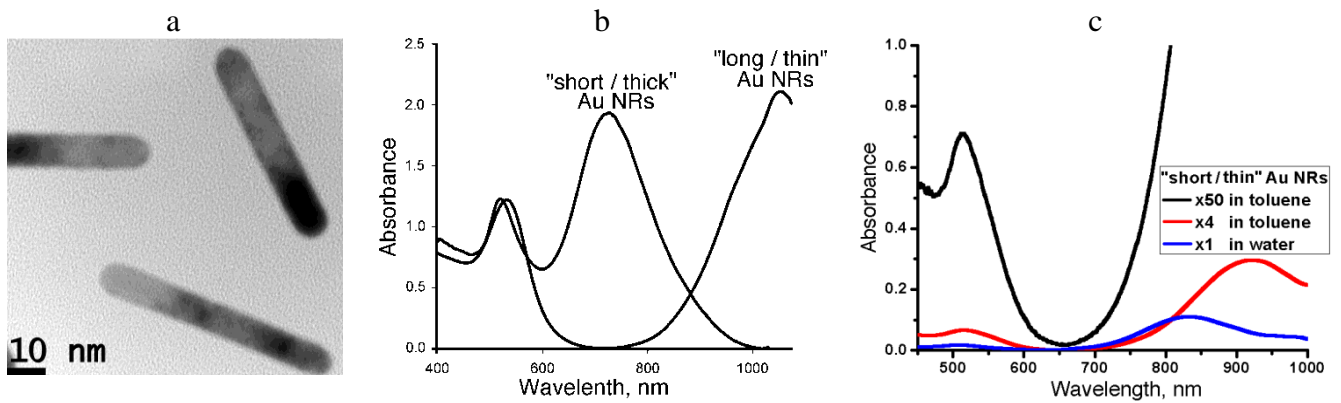

The "short/thick" NRs are suitable to explore spatial distribution and orientation of NRs by monitoring absorption near 725 nm, while the "long/thin" NRs are better suited to observe the cloaking effect. The results were similar for water and toluene dispersions. However, since the refractive index of water does not match that of glass capillaries confining the dispersions, we describe only the toluene case.

## 2.2. Two Types of Samples: Flat and Cylindrical Cells

We study two different geometries, flat cells and cylindrical cells. (1) The flat cells are formed in between two glass plates, with two mutually perpendicular electrodes in the plane of the cell, Fig.2 (a,b). One (grounded) electrode is a copper wire of diameter 2 μm in a borosilicate glass shell of diameter 20 μm that determines the separation between the glass plates. The second electrode is a similar wire (with the glass shell stripped near the tip) connected to a waveform generator. The cell is filled with the toluene dispersion of Au NRs and sealed. The gradient electric field $\mathbf{E}_e$ in the crossed geometry of electrodes in the flat cell mimics the radial gradient in the cylindrical sample, Fig.2 (c). (2) The cylindrical sample represents a circular capillary. The electric field $\mathbf{E}_e$ is created by coaxial electrodes; one is a bare copper wire of diameter 2 μm running along the axis and the second one is a transparent layer of indium tin oxide (ITO) deposited at the outer surface of the capillary. The space between the inner surface of glass capillary and the central electrode is filled with the dispersion of NRs that represents our electrically controlled metamaterial shell. The central electrode (2) plays a dual role, setting up the gradient electric field and also serving as the object to be "cloaked" by the shell.

**Figure 2.** Samples used in the experiments: The flat (**a,b**) sample formed between two glass plates (1) with orthogonal copper wires (2) and (3), filled with Au NRs dispersed in toluene (4) that are isotropically distributed when the electric field is off (**a**) and form a condensed oriented structure when the field is on (**b**). The cylindrical sample (**c**) in a glass capillary (1) with coaxial electrodes (2) and (3); the cavity is filled with Au NRs dispersed in toluene (4) and sealed by a transparent optical adhesive (5).

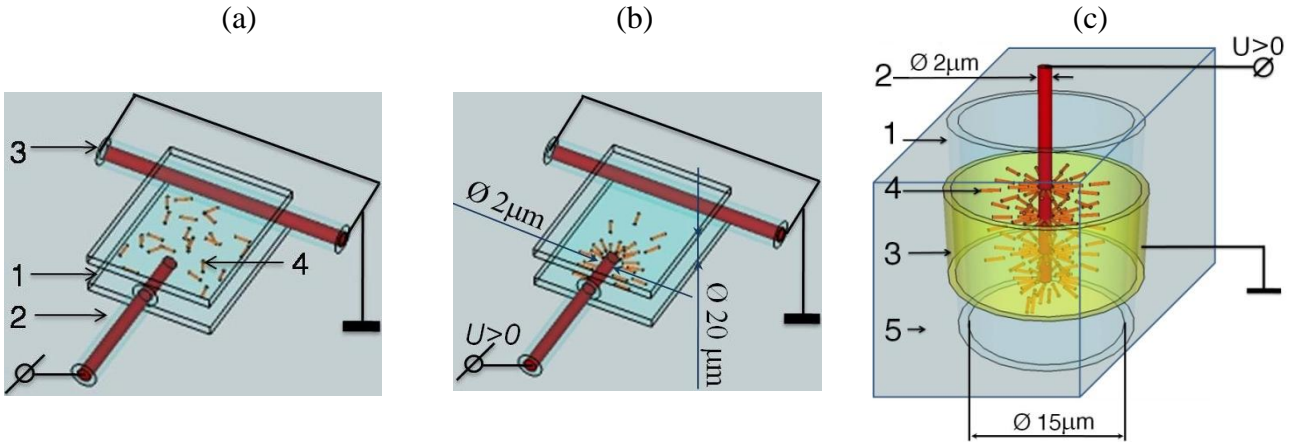

## 2.3. Dielectrophoretic Force

The size of NRs is much smaller than the characteristic scale of the electric field gradients, thus the dielectrophoretic force $\mathrm{F}_{DEP}$ acting on a NR can be calculated in dipole approximation [20] $\boldsymbol{F}_{DEP} = -\boldsymbol{\nabla}W$, where the potential $W$ is

$$W = -\frac{1}{2}Re[\alpha^*]VE_{e,rms}^2, \tag{1}$$

$V$ is the volume of the NR, $E_{e,rms}$ is the root-mean-square (*rms*) value of the electric field, and $Re[\alpha^*]$ is the real part of the effective complex polarizability written for an elongated particle as [20]:

$$\alpha^* = \varepsilon_t \cdot \frac{\varepsilon_{NR}^* - \varepsilon_t^*}{\varepsilon_t^* + A_d(\varepsilon_{NR}^* - \varepsilon_t^*)} \qquad . \tag{2}$$

Here $\varepsilon^* = \varepsilon - i\sigma/\omega$ is the complex permittivity of NRs and the medium (subscripts "NR" and "t", respectively), $\sigma$ is the conductivity, $\omega = 2\pi f$, $A_d$ is the depolarization factor that depends on the orientation of the NR with respect to the electric field. With $\varepsilon_t = 2.4\varepsilon_0$, $\varepsilon_0 = 8.854 \times 10^{-12}\ C/(Vm)$, $\varepsilon_{NR} = -6.9\varepsilon_0$, $\sigma_{NR} = 4.5 \times 10^7\ S/m$, $\sigma_t \sim 5 \times 10^{-11}\ S/m$, $f = 10^5\ Hz$, one finds $|\varepsilon_{NR}^*/\varepsilon t*\sim 10^{13}$ so that the expression for the real part of the effective complex polarizability simplifies to

$$\alpha^* = \frac{\varepsilon_t}{A_d}. \tag{3}$$

Using the typical NR volume $V = \pi d_{NR}^2 L_{NR}/4 \sim 2 \times 10^{-23} m^3$, applied field $E_e \sim 10^7\ V/m$, and the scale of gradient $l \sim 10\ \mu m$, one estimates the dielectrophoretic force acting on an isolated NR of a modest aspect ratio yielding a depolarization factor $A_d = 0.2$, as $F_{DEP} \sim 10\ fN$. The corresponding potential $W \sim F_{DEP} l \sim 10^{-19}$ J $\sim 10 k_B T$ ($T$ is the room temperature) is high enough to overcome the Brownian randomization and to accumulate the NRs in the regions of maximum field. This estimate also suggests that the major axes of NRs (corresponding to the smallest depolarization factor $A_d$) orient along the field and that the medium becomes structurally and optically similar to a uniaxial nematic liquid crystals, with NRs being the building units.

The depolarization factor for the major axis of a NR can be calculated by using a model of prolate spheroid with axes $a_1 > a_2 = a_3$ and eccentricity $g = \sqrt{1 - a_3^2/a_1^2}$, see, e.g. [31]:

$$A_{d1} = \frac{1-g^2}{2g^3}\left(\ln\frac{1+g}{1-g} - 2g\right). \tag{4}$$

For a spheroid with the aspect ratio $a_1/a_3 = 4$, one finds $A_{d1} = 0.075$. Numerical simulations show that the difference in the depolarization factors calculated for cylinders and spheroids of the same aspect ratio is small, less than 5% [31, 32].

We use a commercial Finite Element Package of COMSOL Multiphysics with AC/DC module, version 4.0a, to simulate the electric field patterns and dielectrophoretic potentials, Eqs.(1,3,4), in the flat and cylindrical cells, for $l_{NR}/d_{NR} = 4.5$, Fig.3. The geometry (diameter of electrodes, distance between them) and material properties chosen for simulations are close to the experimental parameters. Numerical simulations show that the dielectrophoretic potentials $W$ in the cylindrical and flat cells are similar to each other. The flat cells thus represent a convenient experimental model of the cylindrical cell, mimicking the cross-section of the latter which is hard to visualize in real experiments. Note, however, that for the flat cell, the simulations are 2D and do not take into account field variations along the coordinate normal to the cell, which is an oversimplification of a real experimental situation.

**Figure 3.** Spatial map of the electric field in the cylindrical cell filled with pure toluene, under an applied voltage 200V; 1 is the glass capillary, 2 is the central electrodes running along the axis of the cylindrical cavity, 3 is the outer electrode, and 4 is toluene, filling the gap between the central electrode and the inner surface of the glass shell (**a**). The same for the flat cell; 2 is the "central" electrode connected to a waveform generator, 3 is the grounded electrode perpendicular to the electrode 2, and 4 is toluene (**b**). The electric field (**c**) and the dielectrophoretic potential (**d**) for two cells, as the function of a radial distance $r$ measured from the axis of the cylindrical cell in (**a**) and from the center of the semispherical tip of central electrode in (**b**).

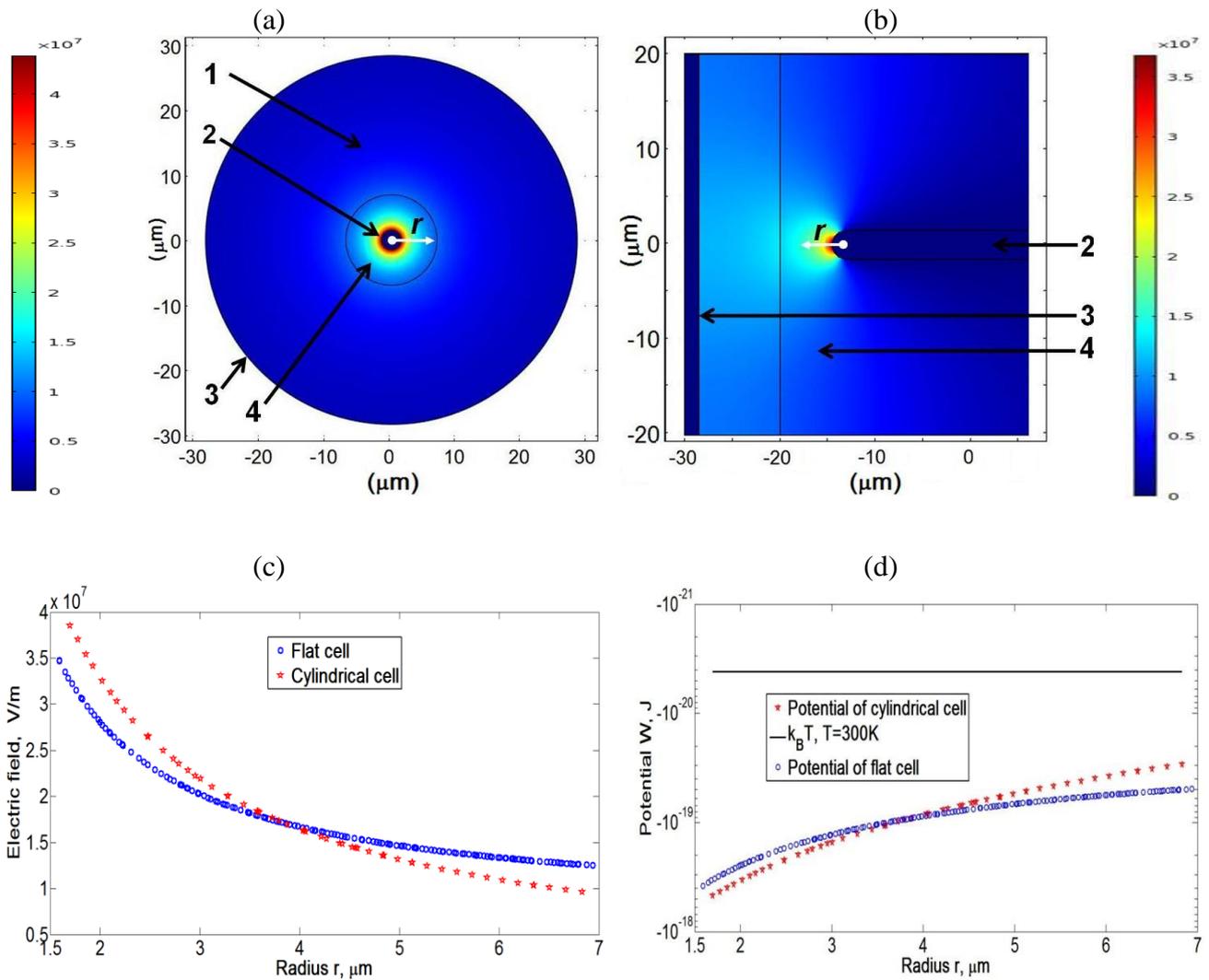

## 3. Experimental Results

The model [12] considers the optical cloaking effect achieved by a metal-dielectric shell of variable composition with the outer and inner diameters $D \approx 3.2 d_O$ and $d_O \approx 2~\mu m$, respectively. To yield a monotonous decrease of the effective refractive index from 1 to 0 between them, the filling factor of metal wires should gradually increase as one moves towards the inner surface. In theory [12], the feat is achieved by using a solid dielectric with metal wires piercing it along the radial directions; the filling factor increases near the inner surface and the whole structure is kept together mechanically.

In our approach, the dielectric is fluid rather than solid and the filling factor is nothing else but the spatially varying volume fraction $\eta(\mathbf{E}_e(\mathbf{r}))$ of NRs. A gradient electric field $\mathbf{E}_e(\mathbf{r})$ is applied to the dispersion of metal NRs in a dielectric fluid to create a dielectrophoretic force that condenses and aligns the NRs in a radial fashion. The resulting spatially varying $\eta(\mathbf{r})$ is determined by the dielectrophoretic coupling with the applied field and also by the forces that oppose it, such as the osmotic pressure, repulsive electrostatic and steric interactions of NRs. One of the important goals of this work is to establish the dependency $\eta(\mathbf{r})$ experimentally. It is expected that the volume fraction is increasing towards the maximum of the field, in our particular examples, towards the central electrode.

### 3.1. Flat Cells

At zero field, the NRs are distributed randomly, Fig.2(a), as their volume fraction in toluene dispersion is orders of magnitude lower than the one needed to form a nematic liquid crystal of the Onsager type (caused by steric repulsions). There is no preferred alignment, and the optical appearance of the cell does not depend on light polarization. When viewed between two crossed polarizers, the cell appears dark. When the AC field $\mathbf{E}_e$ (typical frequency 100 kHz) is applied, the Au NRs, being more polarisable than toluene, move into the regions of high electric field because of the dielectrophoretic effect [20], Fig. 2(b). The flat cell design is convenient for the analysis of field-induced radial gradients of structural and optical properties of the dispersions.

Observations under a microscope with two parallel polarizers reveal that the field accumulates the Au NRs near the central electrode, Fig.4. We quantify the spatial distribution of NRs by measuring the intensity of light transmitted through the cell as a function of a spatial coordinate along the line OX, crossing the central electrode of the flat cell near the tip, Fig.4. The transmission is lower for light polarization parallel to OX than for light polarized perpendicular to it, Fig.4c, suggesting that the NRs are aligned perpendicularly to the central electrode's surface.

**Figure 4.** Voltage-condensed ($U$=200 V$_{rms}$, $f$=100 kHz) cloud of "short/thin" Au NRs in toluene, accumulated near the central electrode of a flat cell. The sample is viewed under a microscope between two parallel polarizers with the transmission direction $\mathbf{E}$ either perpendicular (**a**) or parallel (**b**) to the line OX crossing the central electrode, in the spectral region (550-700) nm of the CCD camera. Light transmission along the line OX for the two polarizations (**c**).

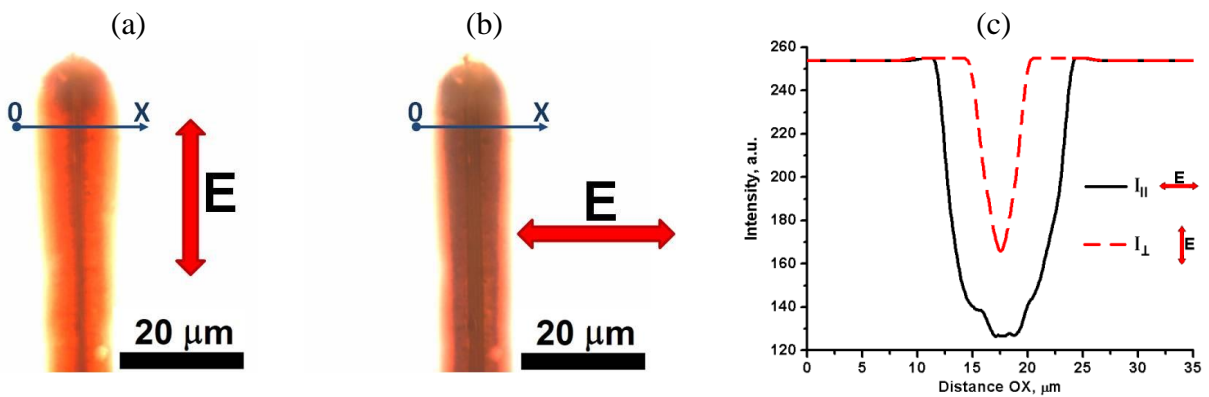

To characterize the concentration gradients of "short/thin" Au NRs, we measured the transmittance profiles of light polarized parallel to OX, as the function of the applied voltage, for the wavelength $\lambda = 460\,nm$, for which the dependence of absorption on the orientation of NRs was found to be relatively weak. Near the central electrode, light transmittance is reduced when the voltage is on, confirming accumulation of NRs, Fig.5(c). Since the absorption depends exponentially on the concentration of absorbing particles and the cell thickness, we determine the ratio $k_{NR}(x) = \eta_U(x)/\eta_O$ as the measure of how much the local field-induced filling factor $\eta_U(x)$ of Au NRs averaged along the cell thickness is larger than the initial (field-free) filling factor $\eta_O$. We estimate $\eta_U(x)$ from the transmittance $lnT_\parallel(x) \sim -\eta_U(x)d_c$, where $d_c = 20\,\mu m$ is the cell thickness. This thickness is too small to determine $\eta_O$ accurately; thus we used transmittance data $A(\lambda) = -log_{10}T$, shown in Fig.1(c) which we obtained for a thick cell $D_c = 500\,\mu m$: $A(\lambda) \sim \eta_0 D_c$. Figure 5(d) shows that for $\lambda = 460\,nm$, the ratio $k_{NR}(x) = -lnT_\parallel(x)D_c/(A(460nm) \cdot d_c \cdot ln10)$ reaches the values of 55 and higher near the electrode (2), which corresponds to $\eta_U \sim 0.02$.

**Figure 5.** Optical microscope textures of the toluene dispersion of "short/ thin" Au NRs in the flat cell when the field $\mathbf{E}_e$ is off **(a)** and on **(b)**; spatial profiles of transmitted light intensities **(c)** and local filling factors ratio $k_{NR}(x) = lnT_\parallel(x)D_c/(A(460nm) \cdot d_c \cdot ln10)$, measured in a monochromatic (460 nm) linearly polarized light **(d)**.

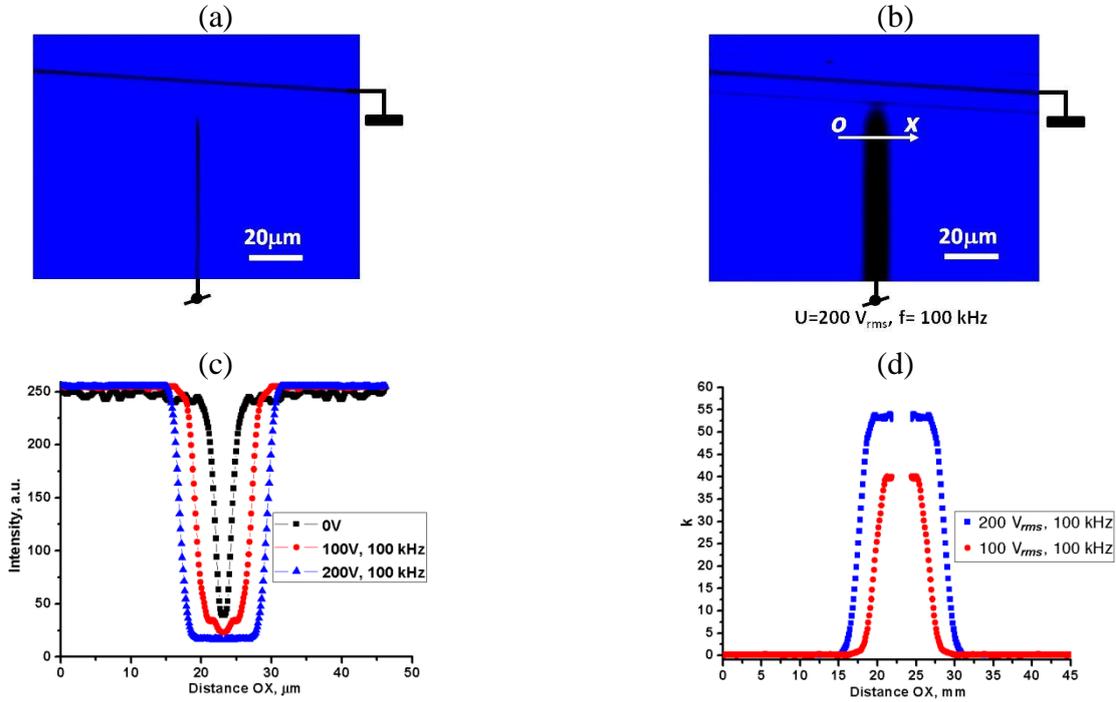

### 3.2. Polarizing Microscopy of Toluene Dispersion of Au NRs in Flat Cells

Under the microscope with crossed polarizers, in absence of $\mathbf{E}_e$, the dispersion of NRs appears dark because it is structurally and optically isotropic. When the field is applied, the clouds of NRs assembled by the field gradients near the central electrode (2), show strong birefringence, which implies an orientational order of NRs, Figs 6,7. The sign of birefringence can be determined with the help of an optical compensator [11]. A waveplate ($\lambda =530$ nm) inserted into the optical pathway of microscope, induces yellow (total retardation less than 530 nm) interference color in the regions where the long axes of NRs are parallel to the slow axis Z' of the waveplate, Figs.6(c,d) and 7(c,d). A blue interference color (retardation higher than 530 nm) is observed in the regions where the NRs are aligned perpendicularly to the slow axis. We conclude that the birefringence of Au NR clouds is negative, i.e., the index of refraction for light polarized parallel to the long axes of Au NRs in dispersion is smaller than for the polarization perpendicular to them.

**Figure 6.** Polarizing microscope textures of the flat cell observed with crossed polarizers A and P. At zero electric field, the toluene dispersion of "short/thin" Au NRs is isotropic and the field of view is dark (**a**). When the voltage is on ($U$=200 V$_{rms}$, $f$=100 kHz), a birefringent cloud of aligned Au NRs appears near the central electrode (2) (**b**). When an optical compensator Z'X', a 530 nm waveplate, is inserted between the sample and the analyzer, yellow and blue interference colors reveal that the field-induced birefringence is negative ((**c**) and (**d**)). Note that reorientation of the sample by 90 degrees from (**c**) to (**d**) causes an interchange of the yellow and blue regions.

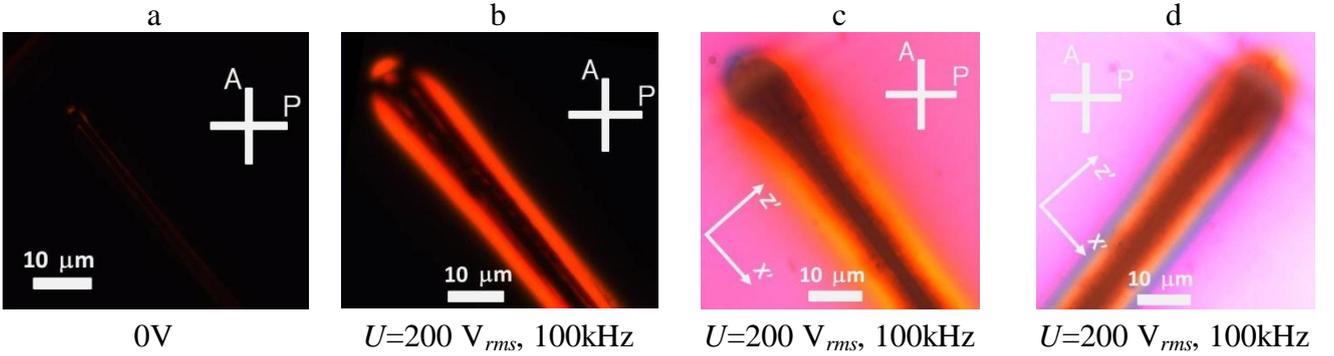

| a | b | c | d |
|---|---|---|---|
| 0V | $U$=200 V$_{rms}$, 100kHz | $U$=200 V$_{rms}$, 100kHz | $U$=200 V$_{rms}$, 100kHz |

**Figure 7.** Polarizing microscope textures of the flat cell filled with "long/thin" Au NRs observed with crossed polarizers A and P (no field in (**a**) and with the field in (**b**)) and with an inserted waveplate ((**c**) and (**d**) with the field on, two different orientations of the sample showing the yellow and blue interference colors interchanged).

(a)          (b)          (c)          (d)

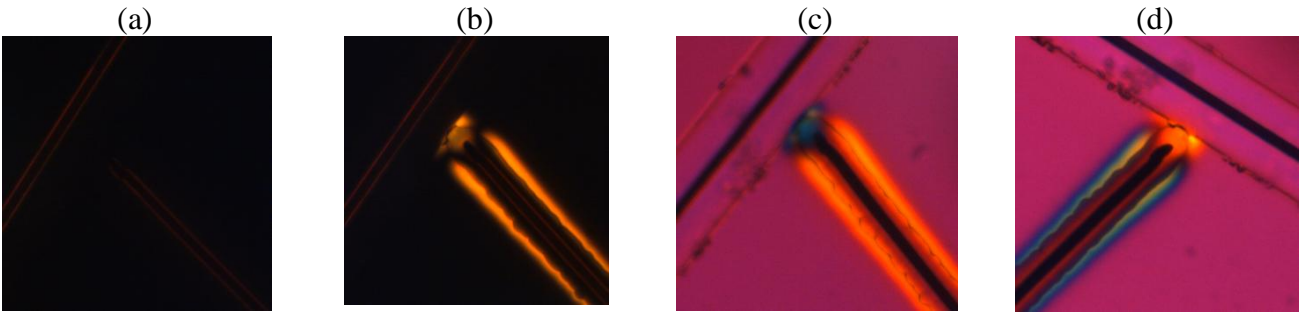

Figure 8 shows the flat cell textures for toluene dispersion of "short/thin" Au NRs viewed between *parallel* polarizers in monochromatic light ($\lambda = 656\,nm$). In zero field, the dispersion is isotropic. When the voltage is applied, the Au NRs accumulate around the central electrode (2). The texture depends on polarization **E** of the probing beam, Fig.8(b, c, d), confirming the optical anisotropy. We determined the spatial profiles of transmitted intensities $I_\parallel^{90}$, $I_\parallel^{0}$, and $I_\parallel^{45}$ along the line OX (Fig.8) that correspond to **E** making an angle 0, 45, and 90 degrees with the central electrode (2), respectively, Fig.9. In Section 4, we will use these profiles to determine the optical path difference between the ordinary and extraordinary waves and to reconstruct the spatial map of optical birefringence. For the same purpose, we determined the light transmission profile $I_\perp^{45}$ for crossed polarizers at the same wavelength 656 nm along the line $O'X'$, Fig.10.

**Figure 8.** Polarizing microscope textures of a flat cell viewed in monochromatic light 656 nm between two *parallel* polarizers, at zero voltage (**a**), at $U = 200$ V$_{rms}$, $f = 100$ kHz (**b**), (**c**), and (**d**). The vector **E** shows the transmission direction of polarizers.

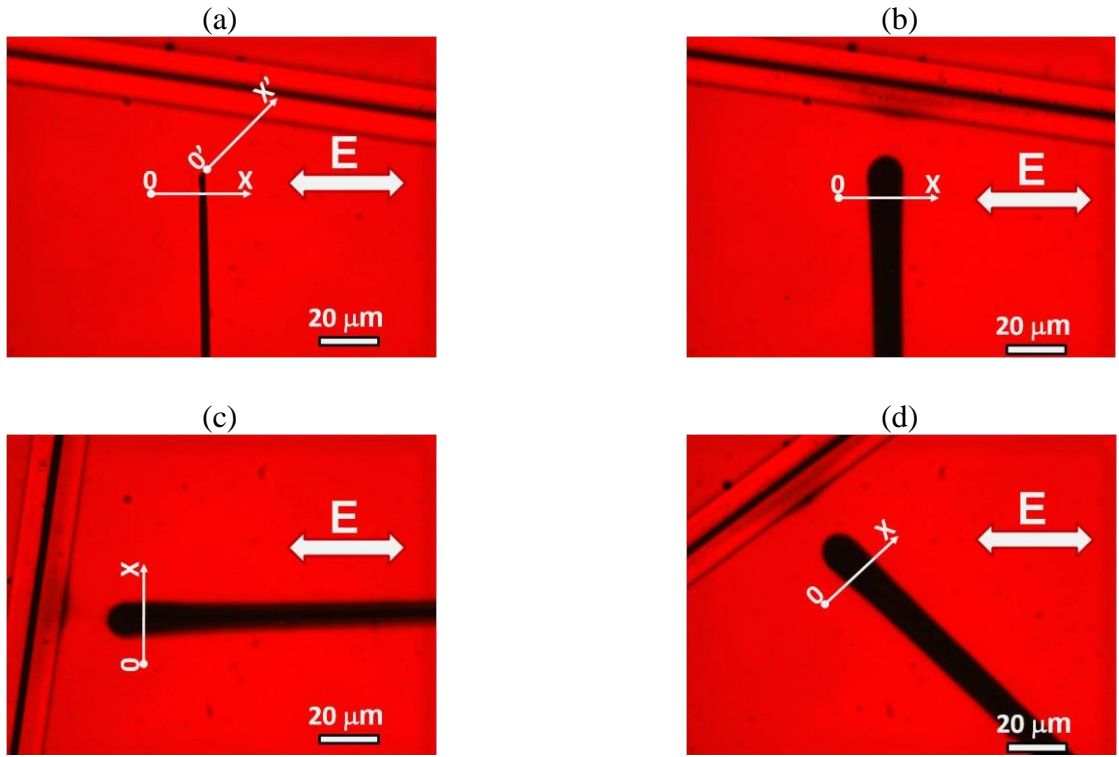

**Figure 9.** Profiles of intensities $I_\parallel^{90}$ (**a**), $I_\parallel^{0}$ (**b**), and $I_\parallel^{45}$ (**c**) vs. distance OX for the flat cell viewed in monochromatic light at 656 nm between parallel polarizers.

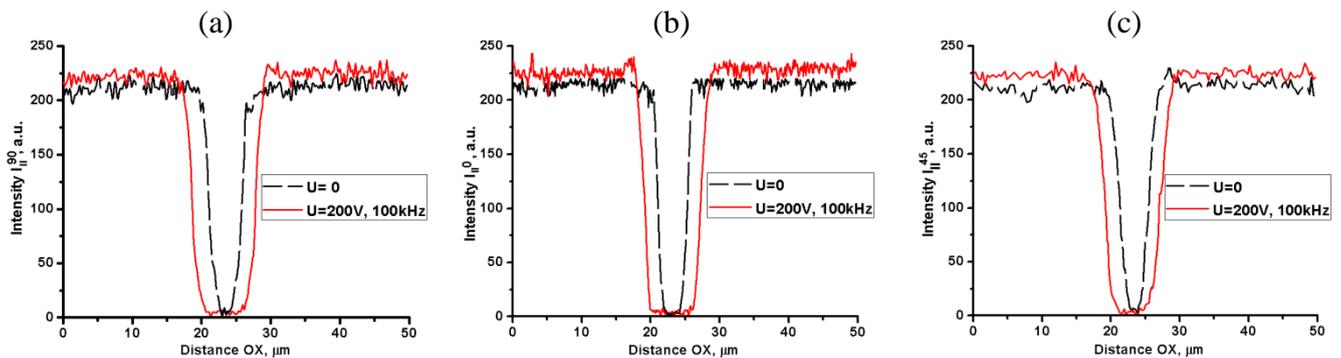

**Figure 10.** Texture of the flat cell with the toluene dispersion of Au NRs under an applied voltage $U$= 200 V$_{rms}$, $f$= 100 kHz, viewed in monochromatic light at 656 nm between two crossed polarizers A and P **(a)**; transmitted light intensity $I_\perp^{45}$ measured along the direction O'X' **(b)**.

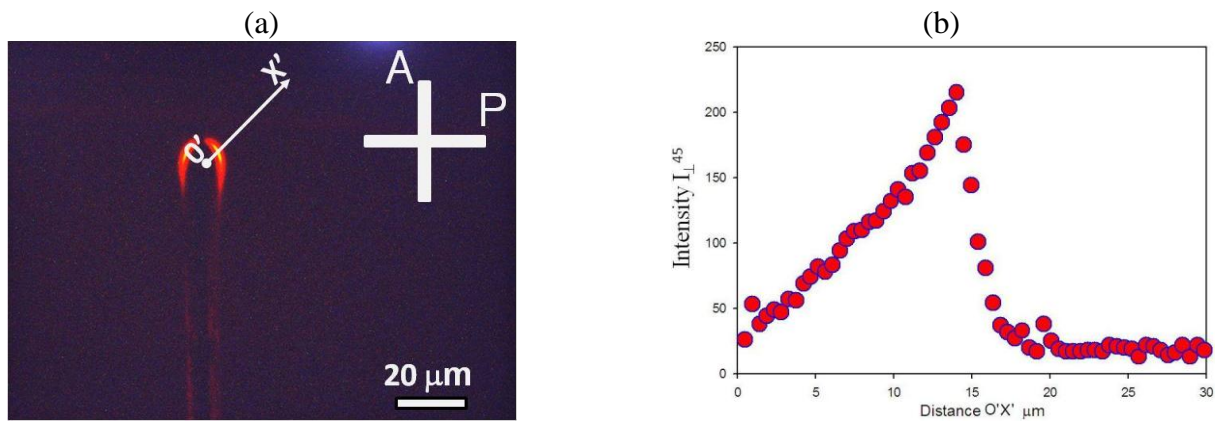

(a)             (b)

### 3.3. Cylindrical Cell: Electrically Controlled Visibility of Central Electrode

The coaxial electrodes in the cylindrical cell create a gradient electric field $E_e \propto 1/r$ that decreases with the distance $r$ from the central electrode, Fig.2(c). Similarly to the case of a flat cell, the AC voltage accumulates and aligns the Au NRs near the central electrode, Figs.11 and 13.

The most striking optical feature of cylindrical cells is that the applied field weakens the shadow of the central electrode, Fig.12, when the latter is observed in the orthoscopic mode under the microscope. The effect is wavelength and polarization dependent, being pronounced for light polarized perpendicularly to the capillary (and thus parallel to the Au NRs), Fig.12(a,b,c). We explored the wavelength dependence for "long/thin" NRs as for these the longitudinal peak of absorption is shifted towards the near infrared region [22]. The transmittance profiles measured for three spectral regions, "red", "green" and "blue" (decoded from the RGB signal of CCD camera) show that the field-induced reduction of shadow is most pronounced in the "red" region with $\lambda = (550 - 700)\ nm$, i.e. where the field-induced birefringence is the highest, Fig.12(a). Propagation of light with parallel polarization is hardly affected by the electric field, Fig.12(d,e,f).

**Figure 11.** The cylindrical cell formed by a glass capillary (1), a copper wire electrode along the capillary axis (2), and a transparent electrode at the outer surface (3). The cavity is filled with "long/thin" NRs in toluene (4) and sealed by polymerized optical adhesive (5). Microscope textures (parallel polarizers) of the capillary when the field $\mathbf{E}_e$ is off (**a**) and on, $U = 170\ V_{rms}$, $f = 100$ kHz (**b**).

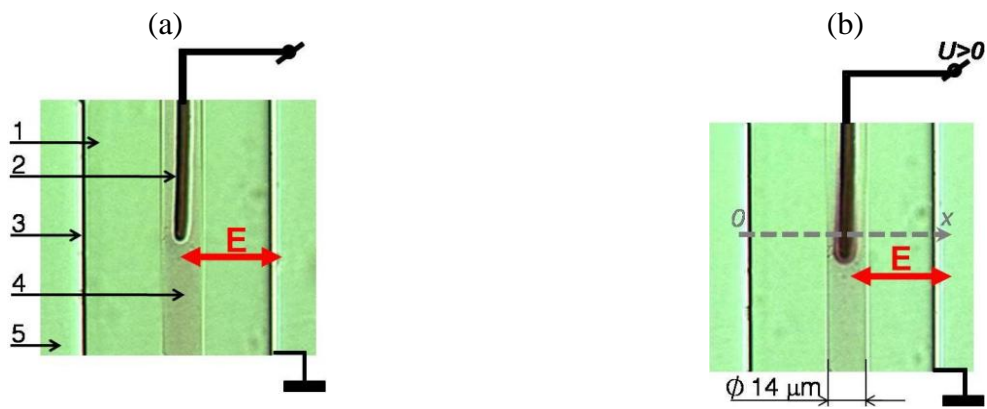

**Figure 12.** Electric field-induced redistribution of "long/thin" Au NRs changes the profiles of light transmission through the capillary for the light polarization perpendicular to the capillary (**a**), (**b**), (**c**), but not for **E** parallel to the capillary (**d**), (**e**), (**f**). Black traces: the field $\mathbf{E}_e$ is off, red traces: field on.

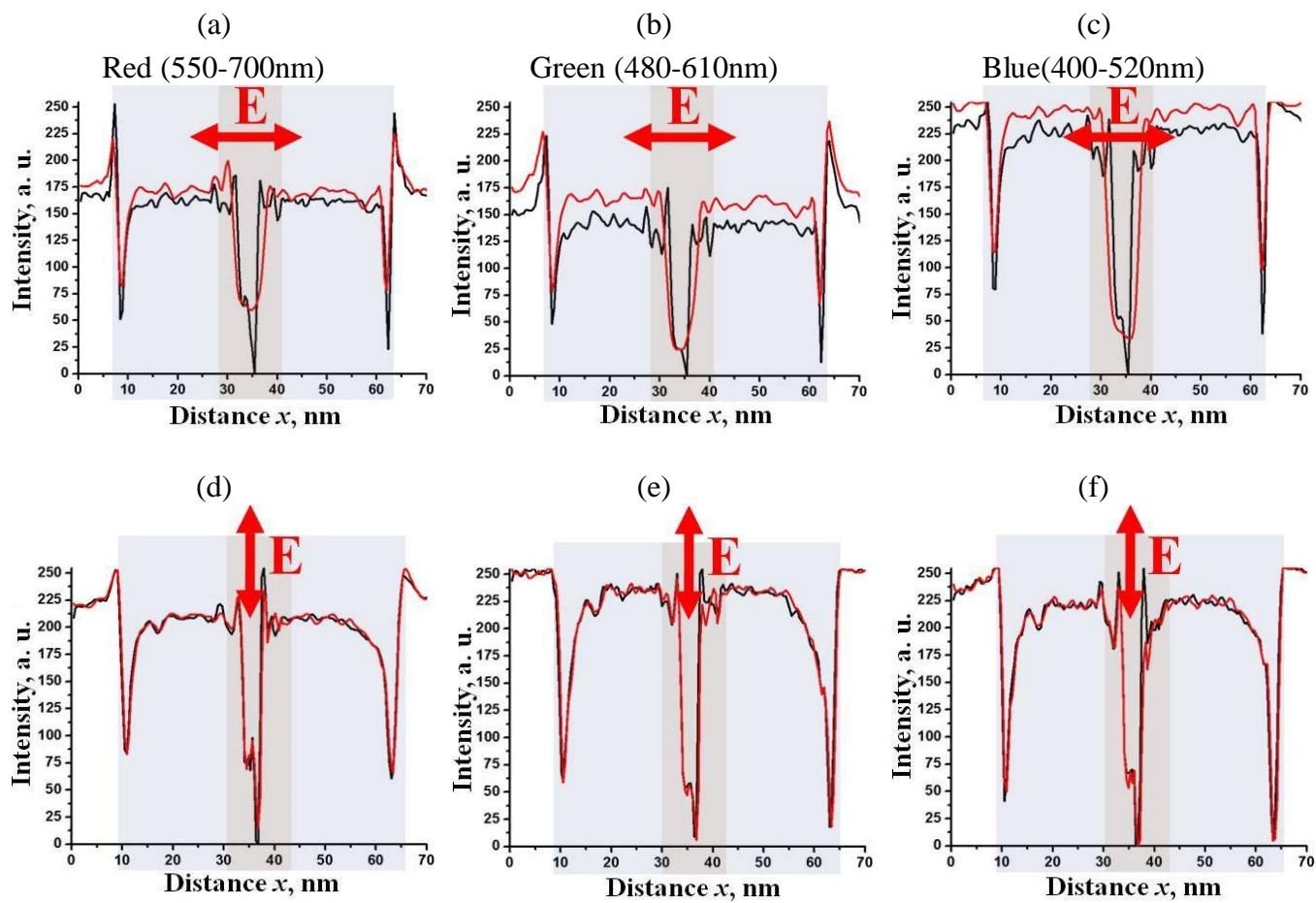

**Figure 13.** Variable visibility of the central electrode (2) in a cylindrical capillary (1) filled with toluene dispersion of "short/thin" Au NRs, shown by the textures at zero voltage (left texture) and at the voltage 90V$_{rms}$, 100 kHz (right texture). Observation under the microscope with light polarized normally to the capillary axis. The right part of the figure illustrates how the light transmission changes along the direction OX for light polarized normally to the capillary (top row) and parallel to it (bottom row).

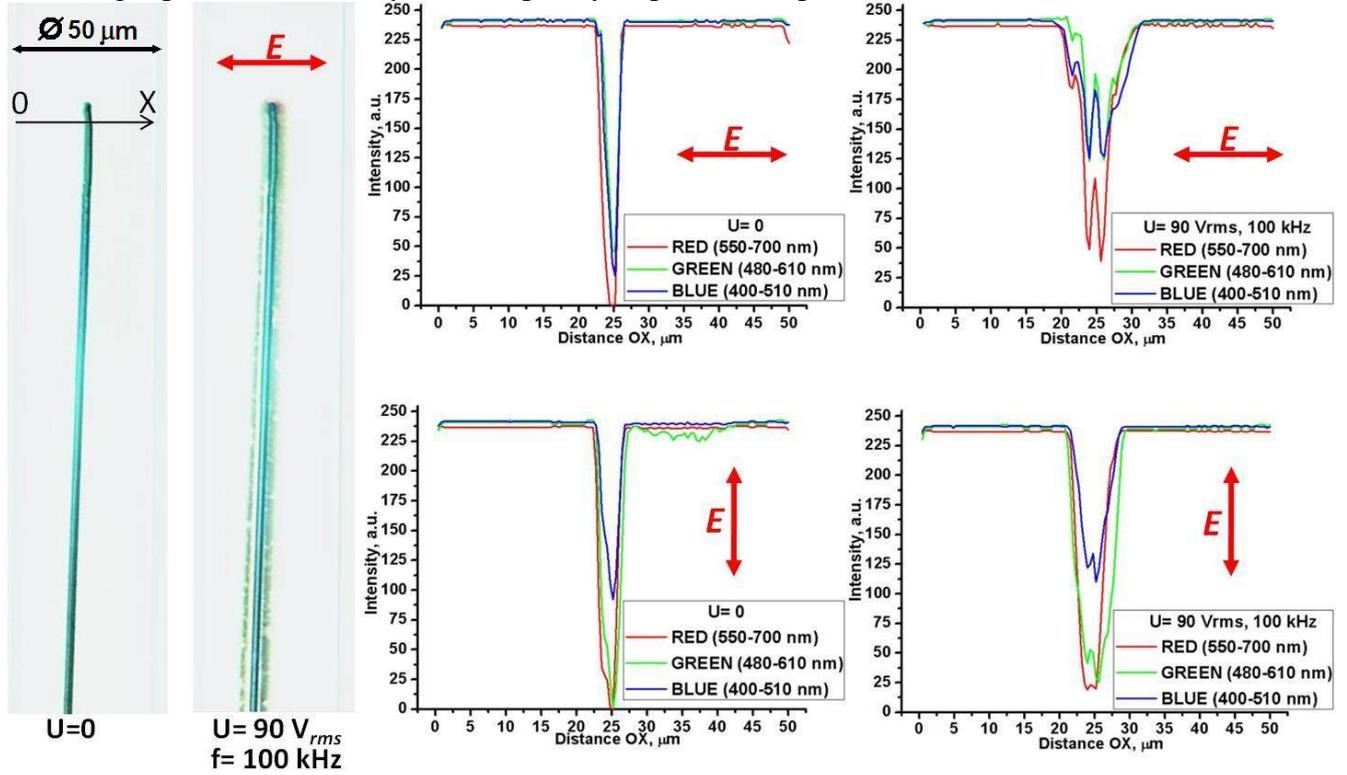

To obtain a better insight into the electric field-induced optical effects, below we analyze the textures theoretically.

## 4. Field Induced Optical Retardation in the Structure of Oriented and Concentrated Au NRs

### 4.1. Light Transmission Through an Absorbing Birefringent Medium

Consider propagation of a linearly polarized monochromatic wave that is normally incident on a slab with unidirectionally aligned NRs. The optic axis is tilted with respect to the slab's normal by an angle $\theta$. The wave splits into the ordinary and extraordinary waves with different indices of refraction $n_o$ and $n_{eff}$, and indices of absorption $\kappa_o$ and $\kappa_{eff}$, respectively. For the ordinary wave, the refractive and absorption indices do not depend on the orientation of the optic axis, *i.e.* $n_o = n_\perp$ and $\kappa_o = \kappa_\perp$, where the subscript $\perp$ means that the quantity was measured with the light polarized normally to the optic axis. For the extraordinary wave, $n_{eff}$ and $\kappa_{eff}$ depend on $\theta$ [36, 37]:

$$n_{eff} = \frac{n_{\parallel} n_{\perp}}{\sqrt{n_{\parallel}^2 \cos^2\theta + n_{\perp}^2 \sin^2\theta}}, \tag{5}$$

$$\kappa_{eff} = \frac{n_{eff}^2}{n_{\perp}^2} \kappa_{\perp} \cos^2\theta + \frac{n_{eff}^2}{n_{\parallel}^2} \kappa_{\parallel} \sin^2\theta, \tag{6}$$

where $n_{\parallel}$ and $\kappa_{\parallel}$ correspond to the case when light propagates perpendicularly to the optic axis ($\theta = \pi/2$) with polarization **E** parallel to the optic axis.

To derive an expression for light transmittance through the slab of thickness $d$, viewed between two arbitrary oriented polarizers, we employ the formalism of Jones matrices. Let us choose a Cartesian coordinate system with the Z-axis directed along the wave vector of light. The polarizer P, slab and analyzer A are perpendicular to the Z-direction. The X–axis is chosen to be along the projection of the optic axis (specified by the long axes of Au NRs) onto the plane normal to Z. The transmission direction of the linear polarizer P is oriented at an angle $\alpha$ with respect to X, while the analyzer's direction A makes an angle $\beta$ with X. The electric field $\mathbf{E}^{ex}$ of the wave exiting the analyzer is related to the incoming electric field **E** through the product of Jones matrices:

$$\mathbf{E}^{ex} = ASP\mathbf{E}, \tag{7}$$

where $A = \begin{pmatrix} \cos^2\beta & \sin\beta\cos\beta \\ \sin\beta\cos\beta & \sin^2\beta \end{pmatrix}$ is the Jones matrix of the analyzer, $S = e^{-\frac{2\pi}{\lambda}(in+\kappa)d} \begin{pmatrix} e^{-\frac{iR+D}{2}} & 0 \\ 0 & e^{\frac{iR+D}{2}} \end{pmatrix}$ is the Jones matrix for the slab with the average refractive index $n = \frac{n_{eff}+n_{\perp}}{2}$ and absorption coefficient $\kappa = \frac{\kappa_{eff}+\kappa_{\perp}}{2}$; $\mathbf{E}^{ex} = \begin{pmatrix} E_x^{ex} \\ E_y^{ex} \end{pmatrix}$ stands for the light wave exiting the analyzer, and $P\mathbf{E} = \begin{pmatrix} \cos\alpha \\ \sin\alpha \end{pmatrix}$ stands for the light wave passed through the polarizer. Note that by the last definition for $P\mathbf{E}$ we effectively normalized the amplitude of the electric field exiting the polarizer by the amplitude $E$ of the incoming electric field. In the definition of $S$, we introduce two new notations: the linear birefringence $R$ and the linear dichroism $D$. For a uniformly aligned slab, $R = \frac{2\pi}{\lambda}\left(n_{eff} - n_{\perp}\right)$ and $D = \frac{2\pi}{\lambda}\left(\kappa_{eff} - \kappa_{\perp}\right)$. For a general case, when the orientation of NRs changes with the coordinate $z$ normal to the slab (and the director experiences splay and bend deformations but not the twist deformations), these quantities are represented by integrals:

$$R = \frac{2\pi}{\lambda} \int_0^d \left[n_{eff}(\theta(z)) - n_{\perp}\right]dz, \quad D = \frac{2\pi}{\lambda} \int_0^d \left[\kappa_{eff}(\theta(z)) - \kappa_{\perp}\right]dz. \tag{8}$$

The light transmittance through the system is $T = \mathbf{E}^{ex}\mathbf{E}^{ex*}$, where the *-symbol denotes a complex conjugate. Using Eq. (7) for arbitrary $\alpha$ and $\beta$, we find:

$$T = e^{-\frac{4\pi}{\lambda}kd}\left\{e^{-D}\cos^2\alpha\,\cos^2\beta + \frac{1}{2}\sin2\alpha\,\sin2\beta\,\cos R + e^D\sin^2\alpha\,\sin^2\beta\right\}. \tag{9}$$

For parallel polarizers ($\beta = \alpha$) the transmission reads

$$T = e^{-\frac{4\pi}{\lambda}kd} \left\{ e^{-D}cos^4\alpha + \frac{1}{2}cosRsin^22\alpha + e^Dsin^4\alpha \right\}. \tag{10}$$

The expression for $T_\parallel$ can be rewritten in terms of the transmittances $T_\parallel^0$, $T_\parallel^{45}$, and $T_\parallel^{90}$ between parallel polarizers, corresponding to three different azimuthal orientations of the director, $\alpha = 0, \alpha = 45^o$, and $\alpha = 90^o$, respectively:

$$T_\parallel = T_\parallel^0 cos^4\alpha + \frac{\sqrt{T_\parallel^0 T_\parallel^{90}}}{2}cosRsin^22\alpha + T_\parallel^{90}sin^4\alpha. \tag{11}$$

Measuring $T_\parallel^0$, $T_\parallel^{45}$, and $T_\parallel^{90}$, one deduces the absorption indices $\kappa_{eff}$ and $\kappa_\perp$, dichroism $\kappa_{eff} - \kappa_\perp$, and birefringence $\Delta n_{eff} = n_{eff} - n_\perp$, using the following relationships:

$$T_\parallel^0 = e^{-\frac{4\pi}{\lambda}\kappa_{eff}d}, T_\parallel^{90} = e^{-\frac{4\pi}{\lambda}\kappa_\perp d}, \tag{12}$$

$$cos\frac{2\pi d\Delta n_{eff}}{\lambda} = \frac{4T_\parallel^{45} - (T_\parallel^0 + T_\parallel^{90})}{2\sqrt{T_\parallel^0 T_\parallel^{90}}}. \tag{13}$$

There is an alternative possibility to measure the phase retardation, by placing the sample between two crossed polarizers, in which case $\beta = \alpha - \frac{\pi}{2}$, and the transmission reads

$$T_\perp = \frac{1}{2}e^{-\frac{4\pi}{\lambda}\kappa d}\{coshD - cos R\}sin^22\alpha. \tag{14}$$

Note that Eq.(14) contains three unknown parameters: $\kappa$, $D$ and $R$. Since we are mostly interested in determination of $R$ which is a measure of the field-induced birefringence in the dispersion of NRs, we need to exclude $\kappa$ and $D$ from the consideration. This can be achieved, for example, by measuring three quantities, namely, transmittance $T_\perp^{45}$ between crossed polarizers, for $\alpha = 45^o$; $T_\parallel^0$ determined with a pair of parallel polarizers, $\beta = \alpha = 0$ and $T_\parallel^{90}$ for $\beta = \alpha = \frac{\pi}{2}$. This is precisely the set of parameters that was measured in the experiments illustrated in Fig.9, 10. Equation (14) can be rewritten as

$$T_\perp = \frac{1}{4}\left\{ T_\parallel^0 + T_\parallel^{90} - 2\sqrt{T_\parallel^0 T_\parallel^{90}}\cos R \right\}sin^2 2\alpha \tag{15}$$

which leads to a straightforward expression to determine the field-induced birefringence $\Delta n_{eff} = n_{eff} - n_\perp$ associated with the effective extraordinary index of refraction $n_{eff}$:

$$cos\frac{2\pi d\Delta n_{eff}}{\lambda} = \frac{(T_\parallel^0 + T_\parallel^{90}) - 4T_\perp^{45}}{2\sqrt{T_\parallel^0 T_\parallel^{90}}}. \tag{16}$$

Equation (16) is similar to Eq.(13), as in both cases, the ratio $\left(T_\parallel^0 + T_\parallel^{90}\right)\Big/\left(2\sqrt{T_\parallel^0 T_\parallel^{90}}\right)$ is nothing else but the ratio of the arithmetic $\overline{T} = \left(T_\parallel^0 + T_\parallel^{90}\right)/2$ and geometric mean $\hat{T} = \sqrt{T_\parallel^0 T_\parallel^{90}}$ for $T_\parallel^0$ and $T_\parallel^{90}$. The only difference is that Eq.(13) uses the quantity $T_\parallel^{45}$, while Eq.(16) deals with the quantity $T_\perp^{45}$; the latter might be more convenient to use as it can be measured more accurately, especially in weakly birefringent cases. We used both approaches to derive the map of spatial profile of the field-induced path difference $\Delta L = \Delta n_{eff} d$ in the flat cells. Figure 14(a) shows the profile of $\Delta L = \Delta n_{eff} d$ along the direction OX in Fig. 8(b,c,d) across the central electrode, calculated using Eq.(13). Figure 14(b) shows variation of $\Delta L$ along the different direction O'X' defined in Fig. 10(a); in mapping $\Delta L$, we used Eq.(16) and the data shown in Fig. 10(b). Both approaches produce similar maps, demonstrating that the maximum field-induced optical path difference is about (-250) nm. The approach based on Eq.(16) produces somewhat smoother features in the region of small path difference, apparently because of the higher accuracy in measuring $T_\perp^{45}$ as compared to $T_\parallel^{45}$.

Within the Au NRs cloud, concentration and orientation of the NRs vary along the Z direction. The local optical quantities such as $n_\parallel(r)$ and $n_\perp(r)$ depend on these two and also on the degree of orientational order of NRs and thus also vary with Z. The simple relationship $\Delta L = \Delta n_{eff} d$ thus produces only a rough estimate of the field-induced birefringence $\Delta n(r) = n_\parallel(r) - n_\perp(r)$. We can neglect the spatial variation of $n_\perp$, as light with polarization perpendicular to the optic axis "sees" only the circular cross-sections of NRs that occupy a relatively small fraction of space [12]. Assuming for a moment that the thickness of a highly concentrated part of the Au NRs cloud is approximately equal to the diameter of electrode $d_e \approx 2\,\mu m$, one can roughly estimate the maximum magnitude of field-induced birefringence in Fig.14(a,b) as $\Delta n_{656nm} = \frac{\Delta L}{d_e} = -250nm/2\,\mu m \approx -0.1$. As we shall see in Section 4.2 below, a more refined approach with numerical simulations of light transmittance through the flat cell produces a similar result.

**Figure 14.** Optical path difference $\Delta L = \Delta n_{eff} d$ vs. distance *OX* calculated using Eq.(13) (**a**) and Eq.(16) (**b**). All data correspond to the toluene dispersion of "short/thin" Au NRs, $\lambda = 656\,nm$, applied voltage $U$= 200 V$_{rms}$, $f$=100 kHz.

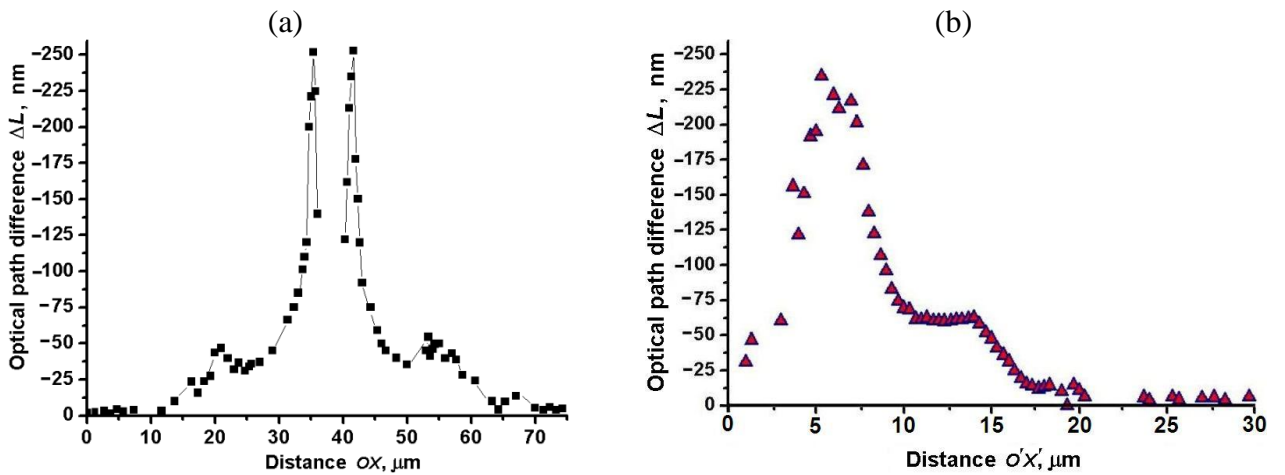

### 4.2. Light transmission through flat samples with NR dispersions

To get a better insight into the magnitude and spatial distribution of the field-induced optical properties of the switchable metamaterial, we need to consider the radial configuration of the optic axis and account for the fact that the system is spatially limited. We calculate light transmission through the sample of thickness $d$, placed between two crossed polarizers. We choose the Cartesian coordinate system $\{x, y, z\}$ with the origin at the wire axis, direct the $z$-axis normal to the substrates, and the $y$-axis along the wire. We assume that the dielectric tensor at optical frequencies is uniaxial with radial ( $r = \sqrt{x^2 + z^2}$ ) dependence of the ordinary $n_\perp(r)$ and extraordinary $n_\parallel(r)$ refractive indices and that the optic axis $\hat{\mathbf{n}}$ is normal to the wire, $\hat{\mathbf{n}} = \{\sin\theta, 0, \cos\theta\}$, where $\tan\theta = x/z$, Fig.15. In such a medium, light propagates along the $z$-axis as an ordinary wave with the refractive index $n_\perp(r)$ and an extraordinary wave with the effective refractive index

$$n_{eff}(r) = \frac{n_\parallel(r) n_\perp(r)}{\sqrt{n_\parallel^2(r)\cos^2\theta + n_\perp^2(r)\sin^2\theta}} \approx n_\perp(r) + \delta(r)\sin^2\theta + \tilde{\delta}(r)\sin^4\theta \qquad (17)$$

where

$$\delta(r) = n_\perp(r)\left[n_\parallel^2(r) - n_\perp^2(r)\right]\Big/2n_\parallel^2(r) \ , \ \ \tilde{\delta}(r) = 3n_\perp(r)\left[n_\parallel^2(r) - n_\perp^2(r)\right]^2\Big/8n_\parallel^4(r) \ . \qquad (18)$$

Here the expansion parameter is $\Delta n(r) = n_\parallel(r) - n_\perp(r)$, because $\delta(r) \approx \Delta n(r)$, and $\tilde{\delta}(r) \approx 3\Delta n^2(r)/4$.

**Figure 15.** Light propagation (vertical green arrow on the right hand side) in a medium with radial configuration of the optic axis (red bars) around the wire (grey circle).

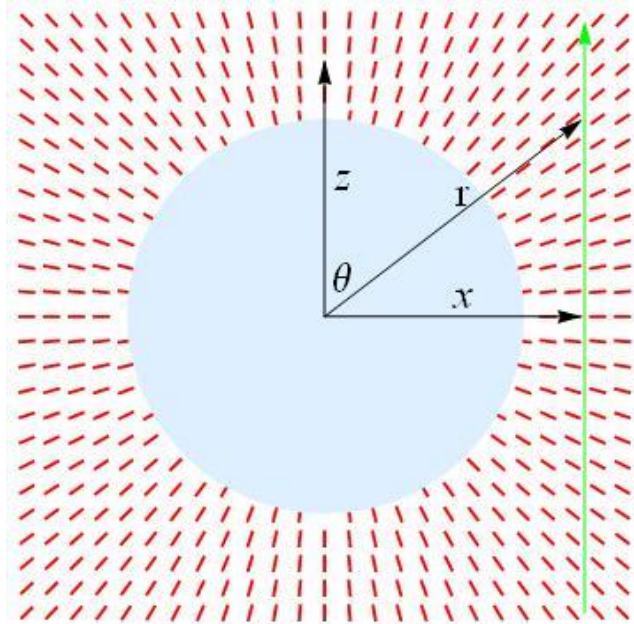

Our goal is to recover $n_\square(r)$ from the experimentally measured optical path difference $\Delta L(x)$ shown in Fig.14(a) (the data in Fig. 14(b) produce similar results). Considering $\Delta n(r) = n_\square(r) - n_\perp(r)$ small, we calculate

$$\Delta L(x) = \int_{-d/2}^{d/2} \left( n_\square(r) - n_\perp(r) \right) dz \approx \int_{-d/2}^{d/2} \delta(r) \sin^2 \theta \, dz \tag{19}$$

We represent $\delta(r)$ as an inverse power series $\delta(r) = \sum_m \delta_m r^{-m}$. Then,

$$\Delta L(x) = x^2 \sum_m \delta_m \int_{-d/2}^{d/2} \left( x^2 + z^2 \right)^{(m+2)/2} dz = d \sum_m \delta_m x^{-m} \, _2F_1\left( 1/2, (m+2)/2, 3/2, -(2x/d)^{-2} \right). \tag{20}$$

Here $_2F_1\left( 1/2, (m+2)/2, 3/2, -(2x/d)^{-2} \right)$ are the Gauss hypergeometric functions, Fig.16, that are linear for small $x$ and saturate to 1 for large $x$,

$$_2F_1\left( 1/2, (m+2)/2, 3/2, -(2x/d)^{-2} \right) = \begin{cases} 1, & x >> d/2 \\ \sqrt{\pi} \, \Gamma\left( (m+1)/2 \right) (2x/d) / 2\Gamma\left( (m+2)/2 \right), & x << d/2, \end{cases} \tag{21}$$

where $\Gamma\left( (m+1)/2 \right)$ is the gamma function.

**Figure 16.** The Gauss hypergeometric functions $_2F_1\left( 1/2, (m+2)/2, 3/2, -(2x/d)^{-2} \right)$ vs. $2x/d$ for different $m$.

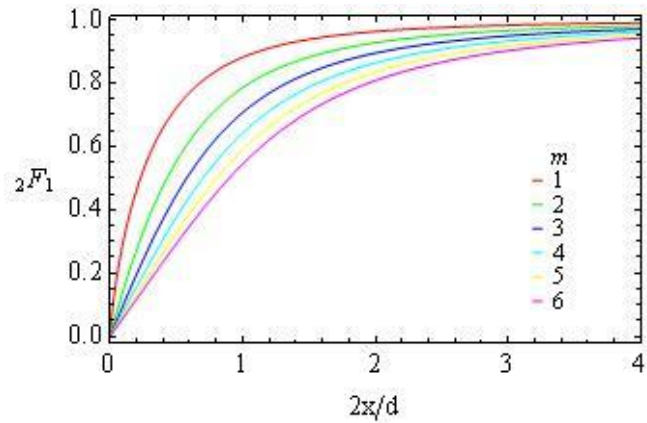

We start the analysis of experimental data with the determination of the center of wire $x_c = 38.29 \pm 0.02 \, \mu m$ by fitting the left wing of the experimental plot in Fig.14(a) with an interpolation from the right wing and vice versa, Fig. 17. Then we combine both wings in Fig.14(a) using $x_c$ as an origin, Fig. 18 and 19, and fit the optical phase retardation profile using different sets of terms in Eq. (20). Figure 18 demonstrates that fitting the left wing (red), right wing (green) and all experimental data (blue) with $m$=3 and $m$=5 from Eq. (20) results in almost the same interpolation curves.

**Figure 17.** Fitting the left wing (red) of Fig. 14 (a) with interpolation from the right wing (green) and vice versa allows us to determine the center of wire $x_c = 38.29 \pm 0.02 \, \mu$m .

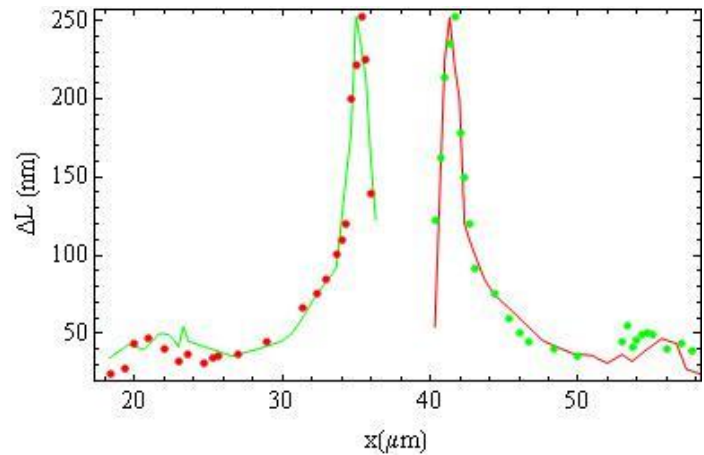

**Figure 18.** Fitting the left part (red), right part (green) and all experimental data (blue) with Eq. (20) with $m$=3 and $m$=5 results in almost the same interpolation curves.

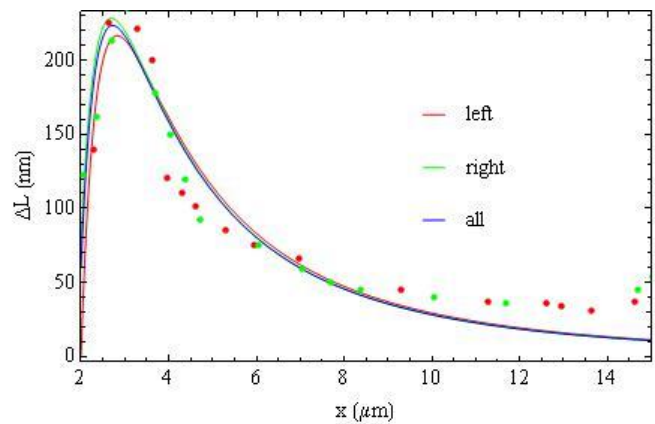

**Figure 19.** Fitting (all) experimental data with Eq. (20) with different sets of $m$, shown in the legend.

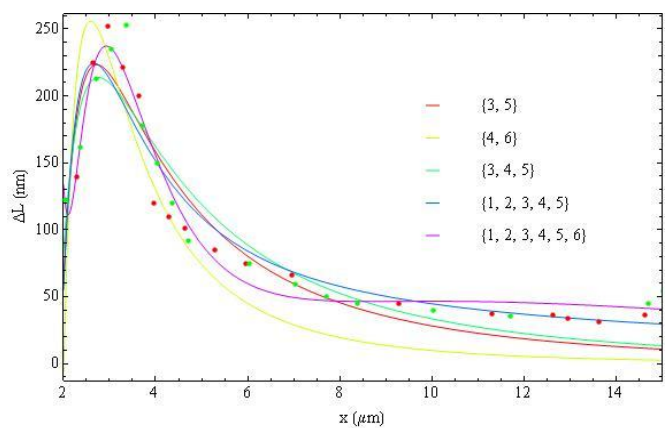

**Figure 20.** Radial dependence of the birefringence parameter $\delta(r)$, Eq. (18), obtained from the fittings shown in Fig.19.

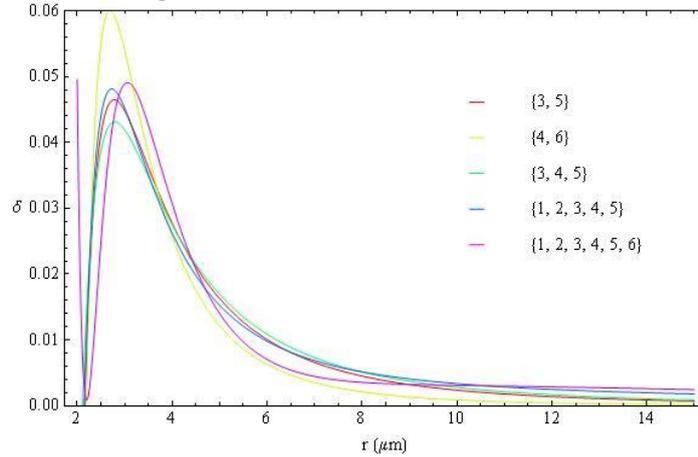

With Fig.19, we produce the fitting of the birefringence parameter $\delta(r)$, Eq. (18), using different sets of $m$, Fig. 20. The fitting curves for different approaches are very close to each other, signaling that the data on $\delta(r)$ are robust. Apparently, the set $m=\{3,5\}$ provides the most reliable fitting because further expansion of the basis of fitting functions does not improve the result substantially. Thus for this set we calculate the radial dependence of permittivity $\varepsilon_r(r) = n_{\parallel}^2(r) = n_{\perp}^3(r)/\left[n_{\perp}(r) - 2\delta(r)\right]$, Fig.21, assuming that the ordinary refractive index $n_{\perp}(r)$ is constant across the capillary and equal to the refractive index of toluene $n_t = 1.49$, so that $\varepsilon_{\theta}(r) = n_{\perp}^2(r) = 2.21$. The spatial distribution of the radial and azimuthal components of dielectric permittivity (refractive indices) induced by the gradient electric field ($U= 200$ V$_{rms}$, $f$=100 kHz) in a flat cell shown in Fig. 21 is one of the major results of this work. These dependences will be used in the numerical simulations of light propagation through the Au NRs dispersion in cylindrical cells, Sect.4.3.

**Figure 21.** Radial dependence of $\varepsilon_r(r) = n_{\parallel}^2(r)$, obtained from Eq. (18) and $\delta(r)$ for $m=\{3,5\}$ shown in Fig.20, with $\varepsilon_{\theta}(r) = n_{\perp}^2(r) = 2.21$. The data correspond to the toluene dispersion of "short/thin" Au NRs in the flat cell, **$\lambda = 656\ nm$**, $U= 200$ V$_{rms}$, $f$=100 kHz, see Fig.14(a).

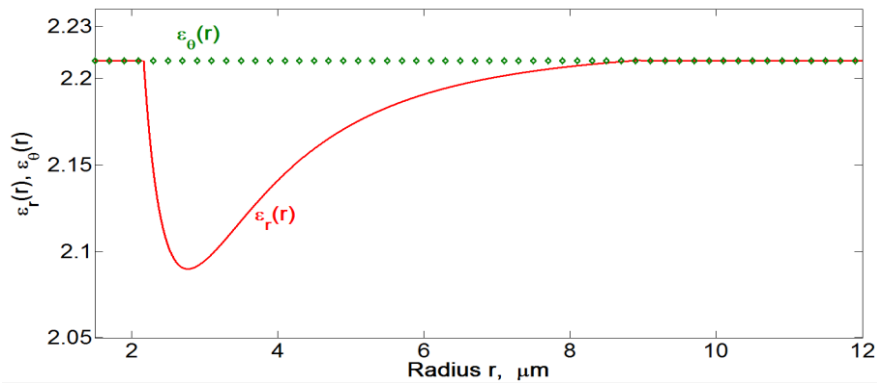

The field-dependent $n_{\parallel}$ can be estimated independently by considering the dispersion as a dielectric of permittivity $\varepsilon_t$ pierced with parallel NRs of permittivity $\varepsilon_{NR}$ with the field-dependent volume fraction $\eta_U$:

$$n_{\parallel} = \sqrt{\left(1-\eta_U\right)\varepsilon_t + \eta_U \varepsilon_{NR}} \ . \tag{22}$$

For the experimentally determined (from light absorption data in Sec.3.1) $\eta_0 = 0.02$, and for $\varepsilon_{NR} = -12.5\varepsilon_0$ at 656 nm [34], one finds $n_{\parallel} \approx 1.4$ and thus $\Delta n \approx -0.1$, the same order of magnitude as other estimates above.

Birefringence of the NRs cloud reflects the cumulative effect of the Au NRs and their polystyrene (PS) coatings. The contribution of PS to the refractive index depends on the configuration of polymer chains covalently grafted to the Au NRs. Birefringence of stretched polystyrene is negative with the refractive index along the PS chain being smaller than the refractive index perpendicular to the chain [35]. Therefore, if the PS chains are directed normally to the NR surface, they will diminish the birefringence effect introduced by alignment of Au NRs; parallel arrangement would enhance the effect of NRs. The experimentally measured birefringence of mechanically stretched PS is $\left|\Delta n_{PS}\right| = 0.0006$ at 700 nm [35]. Therefore, if we assume that in flat cells the entire gap $d \approx 20$ μm is filled with such a birefringent PS, the total optical path difference between the extraordinary and ordinary waves $\Delta L$ would be about 12 nm only, i.e. an order of magnitude smaller than the experimental value of $|\Delta L| = 250$ nm. The estimate suggests that the main contribution in the field-induced modification of the optical properties of the metamaterial in question is produced by the Au NRs themselves.

*4.3 Simulations of optical effects caused by NR redistribution in cylindrical samples.*

We use a commercial Finite Element Package of COMSOL Multiphysics with Radio Frequency module version 4.0a to simulate the electromagnetic wave propagation in the cylindrical cell filled with toluene dispersion of "short/thin" Au NRs. In simulations of the "cloak on" regime, Fig.22(a), we used the dielectric permittivity profile shown in Fig.21, around the central copper electrode of the diameter 2 μm. Note that the permittivity profile in Fig.22 was obtained for the flat cell but in Fig.22(a) it is used to simulate the optical performance of the cylindrical cell. This approximation is justified by the fact that the geometries of gradient electric fields and the dielectrophoretic potentials in flat and cylindrical cells are similar, Fig. 3. Both parts of Fig.22 show the simulated magnetic-field component of the wave propagating throughout the cylindrical shell; the black trajectories show the power flow.

Figure 22 illustrates that when the electric field creates a cloud of Au NRs around the central electrode, Fig.22(a), the shadow of this electrode is mitigated as compared to the case when the electric field is off and the Au NRs are distributed randomly in the cylindrical cavity, Fig.22(b). The power flow near the electrode is bent towards the middle plane of the figure. The effect is the result of the reduced refractive index $n_{\parallel}$ near the central electrode. Of course, the decrease in $n_{\parallel}$ is modest, about 5% of what is used in the theoretical cloak [12], so that the cloaking effect is far from being perfect.

However, the very fact that the electric field gradients are capable of aligning and condensing Au NRs to the extent that the system acquires easily detectable optical anisotropy and variation of the refractive index, is very encouraging for future developments of reconfigurable and switchable optical metamaterials based on dielectrophoretic effects in dispersions of NRs.

**Figure 22**. Simulated light propagation in a cylindrical cell with TM illumination at $\lambda = 656\,nm$. The applied voltage 200 V$_{rms}$ induces the radial profile of the extraordinary refractive index around central electrode and bends the trajectories of power flow around the electrode, mitigating its shadow **(a)**. At zero voltage, the cell has a spatially uniform refractive index (of toluene) and the electrode shadow is well pronounced **(b)**. The color represents the amplitude of magnetic field; see the scale on the right hand side.

(a)

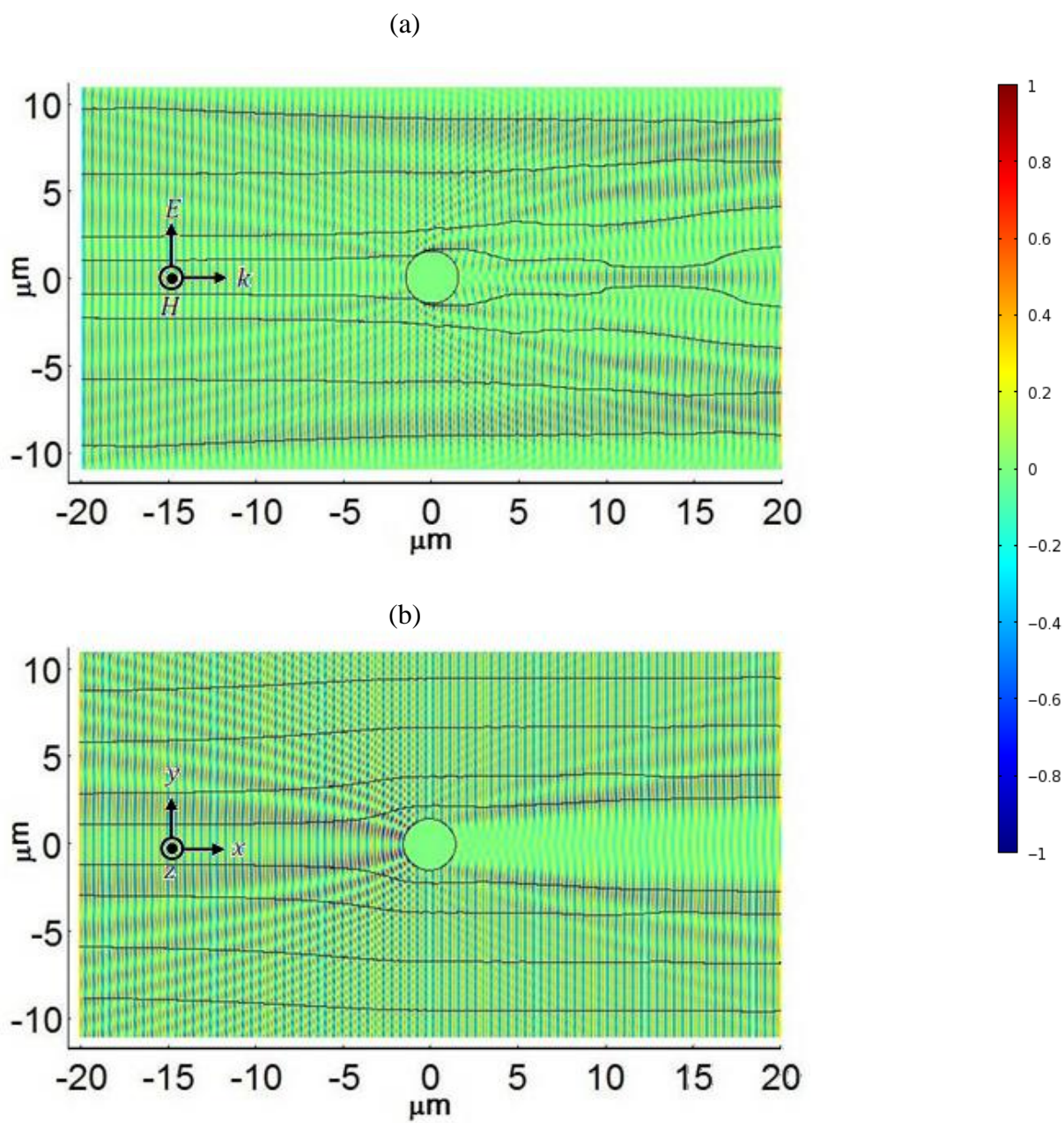

(b)

Figure 23 demonstrates a similar comparison of the "cloak on" and "cloak off" regimes when the light intensity is measured at some distance from the electrode, after the wave travelled to the right, about 19 µm from the electrode. The system parameters are the same as in Fig.22. The plots show transmitted light intensity as a function of the vertical coordinate OY. All plots are normalized by the intensity of the incident TM plane wave. Note that light intensity in the centre of the expected shadow is much higher when the field is on as compared to the case when the field is off, reflecting the bending effect of the Au NRs clouds near the central electrode on the light trajectories. These simulated intensities are similar to the intensity profiles of the red component of RGB signal measured in the experiments with Au NRs, Fig.13.

**Figure 23.** Normalized intensities of incident TM plane wave before the central electrode (red stars), TM wave behind the central electrode with the electric field-induced variation of the extraordinary refractive index (empty blue dots), and TM wave behind the central electrode when the electric field is switched off (solid black line). In the center of shadow, near OY=0, the light intensity in the "field on" case is higher than in the "field off" case.

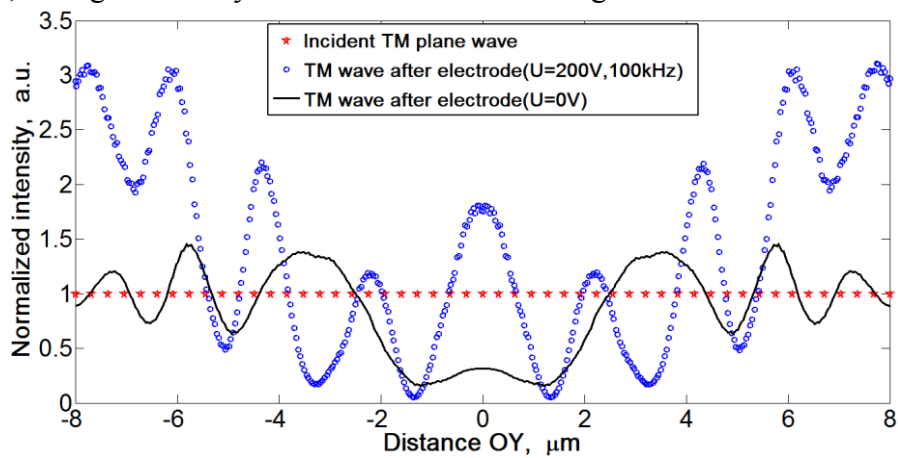

## 4. Conclusions

The experiments above demonstrate that a non-uniform electric field applied to a colloidal dispersion of submicron Au NRs is capable of concentrating the particles in the region of maximum field and also of aligning them parallel to the field lines. This field-induced "liquid crystalline metamaterial" is characterized by a gradient refractive index for polarized light and nonuniform configuration of the optic axis. We thus demonstrate that the approach based on dielectrically controlled dispersions of metal nanorods in dielectric fluids can serve as a broad platform for the development of future complex metamaterial architectures with unique features of electric switching and reconfigurability. In the cylindrical sample, the experiment reproduces the conceptual geometry of the theoretical cloak [12], as the optical axis is directed along the radial directions and the refractive index increases as one moves from the centre of the cylinder to the periphery. The difference is that the experimentally achieved modulation of the refractive index is modest, about 0.05-0.1. Ideally, an efficient metamaterial would have a modulation in the refractive index that is about one order of magnitude higher than the level demonstrated in this work. There are few different ways to enhance the performance. The first factor to improve is the volume fraction $\eta_U$ of the NRs condensed by the gradient electric field. Our experiments reached $\eta_U = 0.02$. To obtain $n_\perp = 0$, according to Eq.(22), one needs to increase $\eta_U$ by one order of magnitude. This appears to be achievable, if one considers the close packing of NRs with not very thick (a few nanometers) aggregation-preventing coatings. Furthermore, the efficiency can be increased by replacing Au with other materials, such as silver (Ag). As shown in ref. [12], for Ag NRs, the filling factor producing a zero refractive index is only 0.125, which is within the reach of the proposed dielectrophoretic approach. Depending on the wavelength of the intended application, other materials might be more efficient, as discussed by Boltasseva and Atwater [38]. The shape of NRs can also be modified to maximize the modulation of the optical properties. For example, as shown by Park et al. [39], the metallic NRs dispersed in dielectric fluids (water) can be reversibly assembled either side-to-side or head-to-head, which would control the position of the plasmonic resonances and increase $\eta_U$. Using a liquid crystal (thermotropic or lyotropic) instead of the isotropic fluid as a dispersive medium can also help in optimizing the proposed reconfigurable metamaterial and enrich the means of structural control.

One of the problems in the development of metamaterials is substantial losses due to absorption. The problem can be addressed by by adding gain materials such as fluorescent dyes [40,41]. This approach should be fully compatible with the proposed metamaterial, as the fluorescent dyes are solvable in dielectric fluids, either water-like, or oil-like.

The main attractive feature of the proposed approach to use metal nanoparticles in dielectric fluids subject to the gradient electric field is in the opportunity to control the optical properties from point to point in space and time. We considered only a radial configuration of the AC electric field. A dielectrophoretic force can also be created in other electrode geometries [27] and by variations in the field phase [20]. It would be of interest to supplement the dielectrophoretic mechanism with effects such as electrophoresis [20]. The electrophoretic force depends on the electric charge on the nanoparticle and is typically linear in the magnitude of the field [20]. A specific case of the electrophoretic effect, called an ''induced charge electrophoresis' [42], is also known for non-symmetric particles. All these mechanisms should add new dimensions to the proposed reconfigurable metamaterials, as they would allow one a better control of nanoparticles. Note that the radial pattern of NRs described in this work is not the only one of interest. For example, simply reversing the concentration gradient of NRs in radial geometry would allow one to switch the metamaterial from the "cloaking" regime to "optical black hole" collector of light [4]. A possibility of such a switching can be explored by combining dielectrophoretic and electrophoretic forces of different direction. These studies are currently in progress.

**Acknowledgements**

This work was supported by AFOSR FA9550-10-1-0527, DOE DE-FG02-06ER46331, and AFOSR MURI FA9550-06-1-0337 grants. We thank N.A. Kotov and P. Palffy-Muhoray for providing us with Au NRs dispersions; A. Agarwal, J. Fontana, P. Luchette, B. Senyuk, H. Wonderly, and L. Qiu for help in sample preparations. We thank P. Palffy-Muhoray, V. M. Shalaev, C. Y. Lee, A. V. Kildishev, and V. P. Drachev for fruitful discussions.